\documentclass[prc,showpacs,nofootinbib,amsmath]{revtex4}
\usepackage{graphics}
\usepackage{epsfig}
\def\lb{\langle}
\def\rb{\rangle}
\def\rhob{{\boldsymbol\rho}}
\def\kappab{{\boldsymbol\kappa}}
\def\calRb{{\boldsymbol{\cal R}}}

\begin{document}

\title{Effective shell model Hamiltonians from density functional theory: quadrupolar and pairing correlations}

\author{R. Rodr\'{\i}guez-Guzm\'an}
\author{Y. Alhassid}
\affiliation{Center for Theoretical Physics, Sloane Physics Laboratory, Yale
University, New Haven, Connecticut 06520, USA}
\author{G.F. Bertsch}
\affiliation{Department of Physics and Institute of Nuclear Theory, BOX 351560, University of Washington, Seattle, Washington
98915, USA}
\date{September 4, 2007}

\begin{abstract}
We describe a procedure for mapping a self-consistent mean-field theory (also known as density functional theory) into a shell model Hamiltonian that includes quadrupole-quadrupole and monopole pairing interactions in a truncated space. We test our method in the deformed $N=Z$ $sd$-shell nuclei $^{20}$Ne, $^{24}$Mg and $^{36}$Ar, starting from the Hartree-Fock plus BCS approximation of the USD shell model interaction.  A similar procedure is then followed using the SLy4 Skyrme energy density functional in the particle-hole channel plus a zero-range density-dependent force in the pairing channel. Using the ground-state solution of this density functional theory at the Hartree-Fock plus BCS level, an effective shell model Hamiltonian is constructed.  We use this mapped Hamiltonian to extract quadrupolar and pairing correlation energies beyond the mean field approximation. The rescaling of the mass quadrupole operator in the truncated shell model space is found to be almost independent of the coupling strength used in the pairing channel of the underlying mean-field theory.
\end{abstract}


\maketitle

\section{Introduction}\label{introduction}

One of the major challenges in nuclear many-body theory is to develop a microscopic and systematic approach that would account for the most relevant correlations in the description of  nuclear properties.  Most existing methods can be divided into two major classes: the self-consistent mean field (SCMF) approximation~\cite{be03}, also known as density functional theory (DFT), and the configuration-interaction shell model (CISM) approach~\cite{br88}. SCMF theories are often cast in terms of an energy density functional that is minimized to obtain the ground-state solution of the system. There exist parameterizations of this energy density functional that are valid globally through the table of nuclei.  Such parametrizations are usually based on the zero-range Skyrme force~\cite{sk56} or on the Gogny interaction~\cite{go75}.  The CISM approach requires as input a large number of interaction matrix elements in a truncated space. While effective CISM interactions can sometimes be traced back to the bare nucleon-nucleon interaction, it is often necessary to adjust them empirically. Since the CISM method requires a truncation to a finite number of shells, the single-particle energies and interaction matrix elements are specific to the mass region under consideration.  Thus the CISM approach lacks the global validity of the SCMF.

On the other hand, the CISM approach has the advantage that the underlying nuclear wave functions (both ground and excited states) are fully correlated. Complete $0 \hbar \omega$ major shell calculations in $p$-~\cite{co65}, $sd$-~\cite{wi84,br88} and $fp$-shell~\cite{ri91,ma97} nuclei have successfully described many of the observed properties of these nuclei. In contrast, the SCMF approach does not account for all correlations present in the nuclear ground state. Partial correlations in a mean-field approach are often accounted for by the spontaneous symmetry breaking of the ground state (e.g., the breaking of rotational symmetry by a deformed ground state)~\cite{ri80}. Additional correlations beyond the mean-field approximation are included by restoration of this broken symmetry (e.g., by angular momentum projection)~\cite{rayner00}, and by configuration mixing of symmetry-projected states via the generator coordinate method (GCM)~\cite{rayner02,be03a,be04}. The GCM is also useful in providing information on excited states of the nucleus. Angular-momentum and particle-number projected GCM has been applied successfully in global studies of binding energies~\cite{be05,sa05,be06} and of spectroscopic properties of the first excited $2^+$ state~\cite{sa07} in even-even nuclei.

In a recent work~\cite{al06} we initiated an approach that takes advantage of the global validity of the SCMF method and the higher accuracy of the CISM approach. It is based on the idea of mapping an SCMF theory onto an effective CISM Hamiltonian that is defined in a suitably chosen truncated model space. Such an approach would enable us to start from an SCMF theory with a global parametrization and construct effective CISM Hamiltonians in different mass regions. In heavy nuclei, the required truncated shell model spaces are expected to be too large for conventional diagonalization of the CISM Hamiltonian. In such cases, one can apply methods that have been developed to treat very large model spaces, such as the shell model Monte Carlo (SMMC) method~\cite{la93,al94}, the Monte Carlo shell model~\cite{ot99} approach, and direct matrix methods~\cite{ca03}. Alternatively, one can use tractable approximations that go beyond the mean-field approximation such as the RPA~\cite{st02}.

A mean-field approximation, based on a Skyrme force, was used in Ref.~\cite{br98} to determine the single-particle energies of a CISM Hamiltonian. In Ref.~\onlinecite{al06} we carried out the first step of mapping an SCMF theory onto the CISM and constructed an effective Hamiltonian that includes quadrupolar correlations. The map was tested successfully in $N=Z$ $sd$-shell nuclei, using the Hartree-Fock (HF) approximation of the so-called USD Hamiltonian~\cite{wi84} as our starting point for an SCMF theory. The USD Hamiltonian was constructed empirically and is known to give a good description of the low-energy spectroscopy of $sd$-shell nuclei.  Using the mapped CISM Hamiltonian, we found that quadrupolar correlations alone account for more than 50\% of the full correlation energies contained in the USD Hamiltonian. We then applied our mapping procedure to study quadrupolar correlations in $N=Z$ $sd$-shell nuclei starting from the SLy4 parametrization of the Skyrme energy functional as the SCMF theory~\cite{al06}.

Here we explore the SCMF to CISM mapping idea further by including pairing
correlations in the mapped effective CISM Hamiltonian in addition to quadrupolar correlations. The starting point is an SCMF theory that includes pairing correlations. Such a theory can be solved in the HF plus BCS approximation [or more generally in the Hartree-Fock-Bogolyubov (HFB) approach] to provide the ground-state single-particle HF wave functions and energies, as well as the BCS occupation amplitudes. A similar HF+BCS approximation can be used to determine a
spherical single-particle basis. This mean-field output is used
to construct an effective CISM Hamiltonian in a truncated spherical model space that
contains both quadrupolar and monopole pairing correlations. After testing the method
using the HF+BCS approximation of the USD interaction as our initial SCMF theory, we
apply it to deformed $N=Z$  $sd$-shell nuclei in the context of DFT. The latter is
based on the SLy4 parameterization of the Skyrme force in the particle-hole channel~\cite{ch98} plus a zero-range density-dependent force in the pairing channel~\cite{te96}. It is solved in the HF+BCS approximation using the Brussels-Paris computer code {\tt ev8}~\cite{bo05,bo85}. The mapping constructed here could also be useful in global studies of level densities within the SMMC approach~\cite{na97,al03}. Both quadrupolar and pairing correlations play an important role in the microscopic calculations of level densities.

Pairing correlations are natural for spherical nuclei. Here, however, we consider the addition of pairing correlations on top of quadrupolar correlations in the CISM Hamiltonian. We therefore focus on nuclei whose HF+BCS ground state is deformed.  In principle, the method presented here can be extended to spherical nuclei by applying an external quadrupole field, thus effectively deforming the nucleus.

The outline of this paper is as follows. In Sec.~\ref{formalism} we discuss
our mapping procedure when pairing correlations are included in the SCMF theory. The ground state is solved in the HF+BCS approximation and its unique
description requires both a density matrix and a pairing tensor (or an anomalous density matrix). The CISM Hamiltonian is constructed in a truncated spherical space to include both quadrupolar and pairing correlations. In Sec.~\ref{USD}, we test our mapping procedure using the HF+BCS approximation of the USD Hamiltonian as our initial SCMF theory for
deformed $N=Z$ $sd$-shell nuclei ($^{20}$Ne, $^{24}$Mg and $^{36}$Ar). In Sec.~\ref{map-DFT} we map a DFT that is based on the SLy4 Skyrme interaction plus a zero-range density-dependent pairing force onto an effective CISM Hamiltonian for deformed $N=Z$ $sd$ shell nuclei. The correlation energies extracted from the mapped Hamiltonian compare well with the correlation energies obtained with other methods. We conclude with a brief discussion of possible future directions in Sec.~\ref{conclusion}.

\section{Theoretical framework}
\label{formalism}

In this section, we discuss the theoretical framework of mapping an SCMF theory into a CISM theory in the case where the SCMF includes pairing correlations. We assume that the SCMF theory is solved in the HF+BCS approximation~\cite{ri80}. This is for example the
case of Skyrme-type density functional theories in which pairing correlations are considered through the addition of a zero-range density-dependent force in the pairing channel~\cite{bo85,bo05}.

The HF+BCS mean-field theory provides us with a set of single-particle HF orbitals ${\phi}_{k}$ and their corresponding single-particle energies ${\epsilon}_{k}$ and BCS occupation amplitudes $v_k$, as well as the total energy $E_{\rm mf}^{\rm HF+BCS}$ of the HF+BCS ground state. For simplicity we assume that this solution, obtained as the global minimum of an energy density functional, is deformed.\footnote{Spherical nuclei can be treated by adding a constraining quadrupole field, leading effectively to a deformed ground state.}  Using the same SCMF theory (i.e., HF+BCS) we can also find the spherical solution. Our plan is to use this output of the SCMF theory to construct an effective CISM Hamiltonian that is defined within a truncated shell model space. In particular, we will consider an CISM Hamiltonian that contains a quadrupole-quadrupole interaction and a monopole pairing interaction.

The CISM is defined in a spherical basis, and we determine such a basis from the spherical solution of the SCMF theory. A spherical solution could be found by iterations when the initial guess for the density matrix is chosen to be spherically symmetric, and as long as the spherical symmetry is preserved in the iteration procedure. In Ref.~\onlinecite{al06} we used the HF approximation, in which case it was necessary (for an open shell nucleus) to use the uniform filling approximation or add a pairing-like interaction with a fixed gap to preserve spherical symmetry in the next iteration.  Here, the BCS approximation guarantees that the density matrix in the next iteration remains spherical.

The construction of a standard spherical basis with good angular momentum $j$ and projection $m$ was discussed in Sec. IV of Ref.~\onlinecite{al06}. Here we use $\alpha$ as a generic label for the spherical basis, i.e., $\alpha$ includes the
quantum numbers $j,m$ as well as other quantum numbers.\footnote{In general we use lowercase greek letters $\alpha,\alpha',\ldots$ to denote spherical states and lowercase roman letters $k,k',\ldots$ to denote deformed states.}

The transformation matrix between the deformed single-particle basis $k$ and the spherical single-particle basis $\alpha$ is denoted by $U_{k \alpha} \equiv \langle k| \alpha \rangle$. We have
\begin{equation}\label{U-trans}
\hat a^\dagger_\alpha = \sum_k U_{k\alpha} \hat b^\dagger_k \;,
\end{equation}
where $\hat a^\dagger_\alpha$ is the creation operator in a spherical state $\alpha$ and $\hat b^\dagger_k$ is the creation operator in a deformed state $k$.

\subsection{Deformed SCMF solution in the HF+BCS theory}\label{deformed}

The deformed ground state in the HF approximation is a Slater determinant $\Pi_k \hat a^\dagger_k |-\rangle$ which is uniquely described by a density matrix $\rhob$. However, in the HF+BCS approximation, the ground state $\Phi = \Pi_{k>0}(u_k + v_k \hat a^\dagger_k \hat a^\dagger_{\bar k}) |-\rangle$  is defined by the BCS amplitudes $u_k,v_k$ (the product runs over half the number of single-particle states $k > 0$ and $\bar k$ is the state conjugate to $k$). The unique description of this BCS ground state requires both a density matrix $\rhob$ and a pairing tensor (or anomalous density) $\kappab$ \cite{ri80}. In the deformed basis
$\rho_{kk'}= \lb \hat b_{k'}^\dagger \hat b_k \rb$ and
$\kappa_{kk'}= \lb \hat b_{k'} \hat b_k \rb$, where $\lb \ldots \rb$ denotes the
expectation value in the HF+BCS ground state $\Phi$. In the spherical basis, we can similarly define
$\rho_{\alpha \alpha'}= \lb \hat a_{\alpha'}^\dagger \hat a_\alpha \rb$
and $\kappa_{\alpha \alpha'}= \lb \hat a_{\alpha'} \hat a_\alpha \rb$.~\footnote{For simplicity of notation, we use the same symbols to denote the matrices in the deformed and spherical basis and they can be distinguished according to their subscripts.} Using the unitary transformation (\ref{U-trans}) from the deformed to the spherical basis, we have
\begin{subequations}\label{trans}
\begin{eqnarray}
\rho_{\alpha \alpha'} & = & \sum_{kk'} U_{k' \alpha'} \rho_{kk'} U^\ast_{k \alpha} \;, \label{rho-trans}\\
\kappa_{\alpha \alpha'} &  = & \sum_{kk'} U^\ast_{k' \alpha'} \kappa_{kk'} U^\ast_{k \alpha} \;, \label{kappa-trans}
\end{eqnarray}
\end{subequations}
or, equivalently, in a matrix notation
\begin{subequations}\label{trans-mat}
\begin{eqnarray}
\rhob & \to & {\bf U}^\dagger \rhob {\bf U} \;, \label{rho-mat-trans}\\
\kappab & \to & {\bf U}^\dagger \kappab {\bf U}^\ast \;. \label{kappa-mat-trans}
\end{eqnarray}
\end{subequations}
According to (\ref{trans-mat}), $\rhob$ transforms as a linear operator under the unitary transformation $U$ while $\kappab$ transforms as an {\em antilinear} operator.

Just as in the HF approximation, the trace of $\rhob$ must be equal to the total number of particles $N$ of a given type (protons or neutrons)
\begin{equation}\label{trace-rho}
{\rm tr}\; \rhob = N \;.
\end{equation}
In the framework of the HF+BCS approximation, this condition is satisfied by using a suitable chemical potential $\mu$ (for each type of particle).

In the HF approximation, the ground state is a Slater determinant and the corresponding density matrix must be a projector $\rhob^2 = \rhob$ with all eigenvalues either 1 or 0. However, in the HF+BCS ground state the matrix $\rhob$ is no longer a projector and its eigenvalues $v_k^2$ are between 0 and 1. The matrices $\rhob$ and $\kappab$, which characterize uniquely the HF+BCS ground state,  satisfy now the relations
\begin{subequations}\label{new-conditions}
\begin{eqnarray}
\rhob -\rhob^2 & = & \kappab \kappab^\dagger \;;\label{condition-1} \\
\rhob \kappab & = & \kappab\rhob^\ast \label{condition-2}\;.
\end{eqnarray}
\end{subequations}

In the following, we assume that the SCMF theory has time-reversal symmetry so
that the single-particle orbitals come in degenerate time-reversed pairs $|k\rb$ and $|\bar k \rb \equiv T|k\rb$ ($T$ is the time-reversal operator). In the following
 we choose the $k>0$ orbitals to be the orbitals with positive $z$-signature~\cite{bo85}. The set $\{k,\bar k \}$ with $k>0$ spans the complete single-particle space. We also assume that the ground state solution is  axially symmetric, in which case $J_z=m$ is a good quantum number ($z$ is the symmetry axis), and the positive $z$-signature orbitals have $m=1/2,-3/2,5/2,\ldots$. Similarly, the spherical states appear
 in degenerate time-reversed pairs $\{\alpha, \bar\alpha \equiv T\alpha \}$ with $\alpha>0$ denoting the positive $z$-signature states.

 The matrices $\rhob$ and $\kappab$ have the following simple form in the deformed HF basis (for $k,k'>0$)
\begin{subequations}\label{hf-basis}
\begin{eqnarray}
\rho_{k k'} & = & \rho_{\bar k \bar k'}  =  v_k^2 \delta_{k k'}\;,\;\;\;\;\;  \rho_{k \bar k'}   = \rho_{\bar k  k'} = 0 \;; \label{rho-hf}\\
\kappa_{k \bar k'} &  =  & -\kappa_{\bar k k'}= u_k v_k \delta_{k k'}\;,\;\;\;\;\;  \kappa_{k k'} =   \kappa_{\bar k \bar k'} = 0  \;. \label{kappa-hf}
\end{eqnarray}
\end{subequations}
The transformation matrix ${\bf U}$ is real (and thus orthogonal) and the matrices $\rhob$ and $\kappab$ remain real in the spherical basis.  Since the spherical states also have good $z$-signature, the transformation ${\bf U}$ does not mix states with different $z$-signature and for $\alpha >0$ we have $|\alpha \rb=\sum_{k>0} U_{k \alpha} | k \rb$. Applying the time-reversal operator to this relation and using the fact that ${\bf U}$ is real, we find that the time-reversed states transform in exactly the same way $|\bar\alpha \rb=\sum_{k>0} U_{k \alpha} | \bar k \rb$. We therefore have
\begin{eqnarray}
U_{\bar k \alpha} = U_{k \bar\alpha}= 0\;;\;\;\;\; U_{k \alpha}= U_{\bar k \bar\alpha} \;.
\end{eqnarray}
It follows that the density and anomalous density matrices have the block form
\begin{eqnarray} \label{block-form}
  \rhob =
\left( \begin{array} {cc}
       {\bf A} &   0   \\
    0  &   {\bf A }  \\
        \end{array} \right) \;;\;\;\;\;
       \kappab = \left( \begin{array} {cc}
      0 & {\bf B}    \\
    -{\bf B}  &    0   \\
        \end{array} \right) \;,
\end{eqnarray}
where ${\bf A}$ and ${\bf B}$ are real matrices whose dimension is half the total number of orbitals. It follows from Eqs.~(\ref{trans}) and (\ref{hf-basis}) that (for $\alpha,\alpha'>0$)
\begin{subequations}\label{A-B}
\begin{eqnarray}
A_{\alpha \alpha'} & = & \sum_{k>0} U_{k \alpha'} v_k^2 U_{k \alpha} \;, \label{A-matrix}\\
B_{\alpha \alpha'} &  = & \sum_{k>0} U_{k \alpha'} u_k v_k U_{k \alpha} \;. \label{B-matrix}
\end{eqnarray}
\end{subequations}
The matrix $\rhob$ is real symmetric while the matrix $\kappab$ is real antisymmetric. The form (\ref{block-form}) is equivalent to
 \begin{subequations}\label{spherical-basis}
\begin{eqnarray}
\rho_{\alpha \alpha'} & = & \rho_{\bar\alpha \bar\alpha'}  = A_{\alpha\alpha'}\;,\;\;\;\;\;  \rho_{\alpha \bar \alpha'}   = \rho_{\bar \alpha   \alpha'} = 0 \;; \label{rho-spherical}\\
\kappa_{\alpha \bar \alpha'} &  =  & -\kappa_{\bar \alpha \alpha'}= B_{\alpha \alpha'}\;,\;\;\;\;\;  \kappa_{\alpha \alpha'} =   \kappa_{\bar\alpha \bar\alpha'} = 0  \;. \label{kappa-spherical}
\end{eqnarray}
\end{subequations}

It is convenient to define the density and anomalous density {\em operators} by
 \begin{subequations}\label{operator-form}
 \begin{eqnarray}
 \hat \rho & = & \sum_{k>0} \left( |k \rb v_k^2 \lb k | + |\bar k \rb v_k^2 \lb \bar k | \right) \;; \\
 \hat \kappa & = & \sum_{k>0} \left( |k \rb u_k v_k \lb \bar k | - |\bar k \rb u_k v_k \lb k | \right) \;,
 \end{eqnarray}
 \end{subequations}
such that their matrix representation in the corresponding basis gives us the
matrices $\rhob$ and $\kappab$, i.e.,  $\rho_{\alpha \alpha'}= \lb \alpha |\hat \rho | \alpha'\rb$ and  $\kappa_{\alpha \alpha'}= \lb \alpha |\hat \kappa | \alpha'\rb$. Equations (\ref{A-B}) and (\ref{spherical-basis}) follow then immediately from the operator form (\ref{operator-form}).

The basic relations (\ref{new-conditions}) defining an HF+BCS state can be rewritten in terms of the matrices ${\bf A}$ and ${\bf B}$ as
\begin{subequations}\label{A-B-conditions}
\begin{eqnarray}
{\bf A} - {\bf A}^2 & = & {\bf B}^2 \;; \\
{\bf A} {\bf B} & = & {\bf B} {\bf A}\;.
\end{eqnarray}
\end{subequations}

A more compact way of representing the relations (\ref{new-conditions}) or (\ref{A-B-conditions}) is to introduce a generalized density matrix $\calRb$ whose dimension is twice the total number of orbitals~\cite{va61,ri80}

\begin{eqnarray} \label{R-matrix}
\calRb \equiv \left (\begin{array} {cc}
       \lb a^\dagger_{\alpha'} a_{\alpha}\rb &    \lb a_{\alpha'} a_{\alpha}\rb  \\ & \\
        \lb a^\dagger_{\alpha'} a^\dagger_{\alpha}\rb &   \lb a_{\alpha'} a^\dagger_{\alpha}\rb \\
       \end{array}  \right) =
\left (\begin{array} {cc}
       \rhob   &   \kappab  \\
       -\kappab^\ast &  1- \rhob^\ast \\
       \end{array}  \right) =
\left (\begin{array} {cccc}
       {\bf A} & 0 & 0 & {\bf B}  \\
       0 & {\bf A} & -{\bf B} & 0\\
        0 & - {\bf B} \;\;& {\bf I - A}\; & 0 \\
        {\bf B} & 0 & 0 & {\bf I - A} \\
       \end{array}  \right) \;,
\end{eqnarray}
where ${\bf I}$ is the identity matrix in a space whose dimension is half the number of orbitals.
Relations (\ref{new-conditions}) or (\ref{A-B-conditions}) are equivalent to the condition that $\calRb$ is a projector
\begin{eqnarray} \label{R-proj_first}
 \calRb^2 = \calRb \;.
\end{eqnarray}
Thus the eigenvalues of $\calRb$ must be 1 or 0. This condition is formally similar to the
one satisfied by $\rhob$ in the HF approximation except that there are now twice as many eigenvalues. In our case, the matrix $\calRb$ is real symmetric and its eigenvalues are pairwise degenerate.

\subsection{Criteria for choosing the truncated CISM space and the rescaling of one-body observables}\label{truncated-space}

As long as the spherical basis $\alpha$ is complete, {\bf U} is unitary and we
have just generated a different representation of the same ground-state density matrix and
pairing tensor. However, in constructing a shell model space, it is necessary to truncate to a valence subspace. We define such a subspace in terms of a projector $\hat P = \sum^\prime_\alpha |\alpha\rb \lb \alpha|$ where $\sum^\prime$ denotes a sum over a subset of the single-particle orbitals $\alpha$. The density operator $\hat\rho$ and anomalous density operator $\hat\kappa$ are then replaced by projected densities
\begin{eqnarray}
\hat\rho \rightarrow \hat P  \hat\rho \hat P\;;\;\;\;\;
\hat \kappa \rightarrow \hat P  \hat\kappa \hat P\;.
\end{eqnarray}
The corresponding matrices in the truncated spherical space are ${\bf P}  \rhob {\bf P}$ and ${\bf P}  \kappab {\bf P}$.

In choosing a suitable truncated subspace, we require the projected densities to preserve approximately their basic properties in the complete space. In the HF+BCS approximation, the trace of the density matrix is constraint to give the total number of particles [see Eq.~(\ref{trace-rho})], and we require that a similar relation is satisfied approximately by the projected density ${\bf P} \rhob {\bf P}$ with the number of particles replaced by the valence number of particles
\begin{eqnarray} \label{trace-P-rho}
{\rm tr}({\bf P} \rhob  {\bf P}) \approx  N_{\rm valence} \;.
\end{eqnarray}
Another requirement is that the conditions (\ref{new-conditions}) are satisfied approximately by the projected
 densities ${\bf P}\rhob {\bf P}$ and ${\bf P}\kappab {\bf P}$.  Alternatively, we require
 the matrices ${\bf P}  {\bf A}{\bf P}$ and ${\bf P}  {\bf B} {\bf P}$ to approximately
 satisfy (\ref{A-B-conditions}). It is convenient to define the
 projected  $\calRb$ matrix
\begin{eqnarray} \label{R-proj}
{\bf P} \calRb {\bf P} =
\left (\begin{array} {cc}
       {\bf P} \rhob  {\bf P}   & \;  {\bf P} \kappab  {\bf P} \\
       -{\bf P} \kappab  {\bf P}  &\; {\bf P}- {\bf P}\rhob  {\bf P}\\
        \end{array}  \right) \;,
\end{eqnarray}
and require that it satisfies approximately the condition (\ref{R-proj_first}), namely that its eigenvalues are close to either 1 or 0. Formally, this condition is similar to the one we require from the projected density matrix $\rhob$  at the HF level~\cite{al06}, except that the projected $\calRb$ matrix is defined in twice the number of dimensions.

In general, we can calculate the expectation value of a one-body observable
$\hat{\cal O}= \sum_{\alpha \alpha'} {\cal O}_{\alpha \alpha'}
\hat{a}_\alpha^\dagger \hat{a}_{\alpha'}$ in the truncated CISM space using the projected density
\begin{eqnarray} \label{truncated-one-body}
\lb \hat{\cal O} \rb = {\rm Tr} (\hat{\cal O} \hat P \hat\rho \hat P)={\sum_{\alpha \alpha'}}^{'} {\cal O}_{\alpha \alpha'} \rho_{\alpha' \alpha}
={\rm tr} \left({\bf {\cal O}} {\bf P} \rhob {\bf P} \right) \;,
\end{eqnarray}
where the sum is restricted to the truncated spherical space. The ratio between the expectation value of $\hat{\cal O}$ in the full and truncated spaces provides us with simple rescaling factors of one-body operators in the truncated CISM space.

\subsection{Effective CISM Hamiltonian}\label{effective-Hamiltonian}

In this section we construct the CISM Hamiltonian given the HF+BCS ground-state solution. In the self-consistent basis $k$ of the HF+BCS, the HF Hamiltonian is
$h_{{\rm mf}, k k^{'}}={\epsilon}_{k} {\delta}_{k k^{'}}$, where $\epsilon_k$ are the energies of the deformed HF single-particle orbitals. We then proceed as in Ref.~\onlinecite{al06}. The HF Hamiltonian matrix ${\bf h}_{\rm mf}$ is transformed to the spherical space and projected on the truncated CISM space. The projected one-body Hamiltonian is
\begin{eqnarray} \label{mf-hamiltonian}
{\bf h} = {\bf P} {\bf U}^T {\bf h}_{\rm mf} {\bf U} {\bf P} \;.
\end{eqnarray}
Next, we expand ${\bf h}$ in multipoles
${\bf h}= \sum_{K} {\bf h}^{(K)}$ and construct the second
quantized one-body tensor operators $\hat{h}^{(K)}$ of rank $K$ [see Eqs.~(10) and (8) in Ref.~\onlinecite{al06}]. The lowest multipole is the monople $\hat{h}^{(0)}=\sum_\alpha\epsilon_\alpha^{(0)} a^\dagger_\alpha a_\alpha$, which we use to define the spherical single-particle Hamiltonian, while the $K=2$ multipole is a quadrupole operator $\hat{h}^{(2)}$, which we use to define a quadrupole-quadrupole interaction.\footnote{Our studies here are limited to even-even nuclei, for which time-reversal invariance implies even values of $K$.}

We now consider the following effective CISM Hamiltonian
\begin{eqnarray} \label{effective-H}
\hat{H} &=& \hat{h}^{(0)} -\frac{1}{2} g_{Q} :\hat{h}^{(2)}\cdot \hat{h}^{(2)}:
  - g_{P} \left( \hat P^\dagger_p \hat P_p + \hat P^\dagger_n \hat P_n \right)
\end{eqnarray}
where $::$ denotes normal ordering and $\hat P^\dagger$ is the monopole pair creation operator
\begin{equation}\label{pair}
\hat P^\dagger = \sum_{\alpha >0} \hat a^\dagger_{\alpha} \hat a^\dagger_{\bar\alpha} \;,
\end{equation}
defined separately for protons and neutrons. Here $g_Q$ and $g_P$ are, respectively, quadrupole and pairing coupling constants. In comparison with the CISM effective Hamiltonian considered in Ref.~\onlinecite{al06}, Eq.~(\ref{effective-H}) contains an additional monopole pairing interaction. We note that one might consider an isovector monopole pairing (which contains also a proton-neutron component). However, in the DFT applications we discuss in Sec.~\ref{map-DFT}, the resulting HF Hamiltonian does not conserve isospin symmetry (because of the Coulomb interaction) and we only consider here an Hamiltonian of the form (\ref{effective-H}).\footnote{For the USD tests in Sec.~\ref{USD} we found that an isospin-invariant interaction constructed from an isovector monopole pair operator gives only a small correction to the correlation energy when compared with a Hamiltonian of the form (\ref{effective-H}).} Both quadrupole-quadrupole and pairing interactions in (\ref{effective-H})
are attractive (i.e., $g_Q,g_P >0$), leading to an CISM Hamiltonian that has a good Monte Carlo sign in the SMMC approach~\cite{na97}.

The method outlined above determines the effective Hamiltonian (\ref{effective-H}) up
to the two coupling constants $g_{Q}$ and $g_{P}$. In Ref.~\onlinecite{al06} we determine the coupling constant of the quadrupolar interaction by matching the deformation energy of the SCMF theory with the deformation energy of the CISM effective Hamiltonian when the latter is solved in the HF approximation. When the effective Hamiltonian has the form (\ref{effective-H}), it is necessary to match two energy scales in order to determine the two coupling constants $g_{Q}$ and $g_{P}$. One energy scale is the deformation energy, here redefined in the HF+BCS approximation
\begin{eqnarray} \label{deformation-energy}
E_{\rm def}^{\rm HF+BCS}= E_{\rm sph}^{\rm HF+BCS} - E_{\rm mf}^{\rm HF+BCS} \;,
\end{eqnarray}
where $E_{\rm sph}^{\rm HF+BCS}$ is the energy of the spherical HF+BCS solution and $E_{\rm mf}^{\rm HF+BCS}$ is the HF+BCS energy of the deformed ground-state solution. The deformation energy in (\ref{deformation-energy}) is usually smaller than the deformation energy calculated in the HF approximation.

To introduce a second energy scale, we define $E_{\rm pair}$ to be the average interaction in the pairing channel
\begin{equation}
E_{\rm pair} = - \frac{1}{2}{\rm tr} (\kappa \Delta) \;,
\end{equation}
where $\Delta$ is the pairing gap matrix.\footnote{The gap matrix is defined by $\Delta_{l l'}= \frac{1}{2} \sum_{k k'} v^A_{l l', k k'} \kappa_{k k'}$, where $v^A$ is the anti-symmetrized two-body interaction.} The energy $E_{\rm pair}$ can be calculated in both the spherical and deformed solutions and we define the pairing energy excess to be
\begin{equation}\label{delta-E-pair}
\delta E_{\rm pair}^{\rm HF+BCS} = E_{\rm pair,mf}^{\rm HF+BCS} - E_{\rm pair,sph}^{\rm HF+BCS} \;.
\end{equation}
Since $E_{\rm pair}$ is lower (i.e., more negative) in the spherical solution, the energy difference in (\ref{delta-E-pair}) is usually positive.

The two coupling parameters $g_Q$ and $g_P$ are determined by matching simultaneously
the two energy scales (\ref{deformation-energy}) and (\ref{delta-E-pair}) between the SCMF theory and the CISM Hamiltonian when the CISM is treated in the HF+BCS approximation.

In general, the coupling parameters should be renormalized depending on the truncation used for the CISM space. These renormalization effects are included implicitly in our procedure by matching the energy scales calculated in the SCMF approach in the complete space with similar energy scales calculated in the truncated CISM space within the HF+BCS approximation.

An important application of our mapping is in the calculation of correlation energies. The correlation energy  $E_{\rm corr}$ is defined as the difference
between the ground-state energy in the HF+BCS approximation to the CISM Hamiltonian and the fully correlated ground-state energy $E_{\rm gs}$ of the CISM
\begin{eqnarray} \label{E-corr}
E_{\rm corr}= E_{\rm mf}^{\rm HF+BCS}- E_{\rm gs} \;.
\end{eqnarray}

Quadrupolar and pairing correlations compete with each other; the quadrupole-quadrupole interaction favors deformation in which pairing is suppressed, while pairing favors the spherical configuration. If we consider the spherical configuration alone, then the BCS approximation already takes into account some of the gain in pairing correlation energy. However, in the deformed nuclei considered here we effectively gain more in mean-field energy by deformation. In the deformed ground state, BCS correlations are often significantly smaller than in the spherical solution (or even vanish). Nevertheless, The correlation energy described by (\ref{E-corr}) is expected to increase when a pairing term is added to the quadrupolar CISM Hamiltonian. This increase arises from fluctuations of the pairing fields  beyond the BCS approximation. In particular, these fluctuations sample the spherical configuration in which pairing correlations are significant.

Even in spherical nuclei (which we do not treat here) we expect to gain a finite correlation energy in (\ref{E-corr}) when adding a pairing term to the CISM Hamiltonian, i.e., there are pairing correlations beyond those described by the spherical BCS solution. The BCS approximation is valid in the limit when the pairing gap is much larger than the mean single-particle level spacing, and studies of pairing effects in ultra-small metallic grains revealed deviations from BCS theory in the crossover to the fluctuation-dominated regime~\cite{delft01,al07a}. Nuclei belong to this crossover regime, in which the pairing gap is comparable to the mean-level spacing~\cite{al07}.

In the remaining part of this section we further motivate our choice of the second energy scale (\ref{delta-E-pair}). As already noted, pairing effects are weaker in a deformed configuration and could vanish altogether (i.e., $\kappab =0$) in the deformed ground state. For example, in our HF+BCS studies of the USD interaction in Sec.~\ref{map-USD}, we observe such a collapse of the BCS solution in all three deformed $N=Z$ $sd$-shell nuclei.  We then have $\delta E_{\rm pair}^{\rm HF+BCS}= - E_{\rm pair, sph}^{\rm HF+BCS}$, and only the spherical solution is relevant for determining the second energy scale. An alternative second energy scale that could be used in such cases is $E_{\rm sph}^{\rm HF} - E_{\rm sph}^{\rm HF+BCS}$, the decrease in spherical energy in the HF+BCS approximation as compared with the HF approximation. In general
\begin{equation}\label{spherical-pairing}
E_{\rm sph}^{\rm HF} - E_{\rm sph}^{\rm HF+BCS}= - E^{\rm HF+BCS}_{\rm pair, sph} - \delta \langle \hat h^{(0)}\rangle - \frac{1}{2}\delta \,{\rm tr} (\rhob_{\rm sph} v^A \rhob_{\rm sph})\;.
\end{equation}
where $\rhob_{\rm sph}$ is the spherical one-body density and $v^A$ is the anti-symmetrized two-body interaction.  The gain in pairing energy from $- E^{\rm HF+BCS}_{\rm pair, sph} > 0$ is thus canceled in part by the increase $\delta \langle \hat h^{(0)} \rangle$ of the one-body energy and the increase $\frac{1}{2}\delta \,{\rm tr} (\rhob_{\rm sph} v^A \rhob_{\rm sph})$ of the average interaction in the particle-hole channel. Despite this partial cancelation, $- E^{\rm HF+BCS}_{\rm pair, sph}$ remains well correlated with $E_{\rm sph}^{\rm HF} - E_{\rm sph}^{\rm HF+BCS}$ and, as long as the deformed solution is unpaired, we find that both energy scales (\ref{delta-E-pair}) and (\ref{spherical-pairing}) lead to CISM Hamiltonians with similar coupling constants.

However, the deformed ground state can generally have a non-vanishing (though weaker) pairing solution, as we observe in our DFT studies of Sec.~\ref{SCMF-results}. In such cases, we find that matching the energy scale (\ref{spherical-pairing}) results in a correlation energy (\ref{E-corr}) of the CISM ground state that is unphysically large. We therefore adopt (\ref{delta-E-pair}) as the second energy scale to be matched when the deformed solution is paired.  The scale (\ref{delta-E-pair}) also has the obvious advantage that it is defined within a single SCMF theory (i.e., HF+BCS) and does not require the use of two different mean-field theories (i.e., HF and HF+BCS).

\section{Tests for the USD shell model Hamiltonian}\label{USD}

In this Section we test the mapping procedure discussed in Sec.~\ref{formalism} by considering the HF+BCS solution of the USD Hamiltonian~\cite{wi84} as our starting point of an SCMF theory. The USD interaction provides a rather good description of binding energies, spectra and electromagnetic transition intensities of $sd$-shell nuclei~\cite{br88}. In these studies we compare the correlation energies found from the mapped CISM Hamiltonian with the full correlation energies of the USD Hamiltonian. In Sec.~\ref{map-DFT}, we apply the mapping formalism to a DFT that is defined by an SLy4 Skyrme force~\cite{ch98} in the particle-hole channel plus a zero-range density-dependent force in the pairing channel~\cite{te96}.

Here and in Sec.~\ref{map-DFT} we focus on $N=Z$ nuclei, in which isovector quadruple-quadrupole correlations could be ignored. We also restrict our studies to nuclei whose self-consistent HF+BCS  ground state is deformed. In the $sd$ shell, this excludes $^{32}$S whose self-consistent HF+BCS ground state is spherical.  Spherical nuclei can be studied by a similar method once a deformation is induced by a quadrupole constraining field.

We also restrict our studies to nuclei in which the spherical configuration supports a BCS solution (so that $\delta E_{\rm pair}^{\rm HF+BCS} \neq 0$). Otherwise, it is necessary to add a constraining pairing field. In the $sd$ shell, this excludes  $^{28}$Si for which both spherical and deformed solutions are found to have no pairing solution. A similar situation occurs in the framework of the DFT application discussed in Sec.~\ref{map-DFT}. Therefore, the $N=Z$  $sd$-shell nuclei we consider here and in Sec.~\ref{map-DFT} are $^{20}$Ne, $^{24}$Mg and $^{36}$Ar.

\subsection{Mapping the HF+BCS theory of the USD interaction} \label{map-USD}

We have carried out HF+BCS calculations using the USD interaction. The resulting
HF+BCS energies are listed in Table \ref{table1} for both spherical and deformed ground-state
configurations.  We find the ground-state solutions to be prolate for $^{20}$Ne and $^{24}$Mg, and  oblate for $^{36}$Ar, with average mass quadrupole moments $\langle \hat Q\rangle$ given in Table \ref{table1}.
The HF+BCS deformation energies (\ref{deformation-energy}) are
calculated to be $5.64, 7.05$ and $1.27$ MeV for $^{20}$Ne, $^{24}$Mg and $^{36}$Ar, respectively. It is interesting to compare these energies with the corresponding HF
deformation energies~\cite{al06} of $10.92, 11.97$ and $3.82$ MeV for the same nuclei. The reduction in deformation energies originates in a negative $E^{\rm HF+BCS}_{\rm pair,sph}$  that lowers the spherical HF energy.

\begin{table}[h!]
\caption{HF+BCS results for the deformed $N=Z$ $sd$-shell nuclei
$^{20}$Ne, $^{24}$Mg and $^{36}$Ar, using the USD interaction and the effective CISM Hamiltonian (\ref{effective-H}). Energies are in MeV, the average quadrupole moment $ \langle \hat Q \rangle$ is in $b^{2}$ (with $b$ being the oscillator radius), $g_{Q}$ is in MeV/$b^{4}$ and $g_P$ is in MeV.
}
\label{table1}
\vspace*{1 mm}
\begin{tabular}{|c|c|c|c|c|c|c|c|c|c|c|c|}
\hline
\hline
Nucleus  &  Interaction   & $g_Q$ & $g_P$ & $E_{\rm sph}^{\rm HF+BCS}$ & $E_{\rm pair,sph}^{\rm HF+BCS}$ & $E_{\rm mf}^{\rm HF+BCS}$ & $E_{\rm pair,mf}^{\rm HF+BCS}$ & $E_{\rm def}^{\rm HF+BCS}$ & $\delta E_{\rm pair}^{\rm HF+BCS}$ & $ \langle \hat Q \rangle$ & $E_{\rm corr}$\\
\hline
$^{20}$Ne  &  USD                &  -      &  -      & -30.74             &   -7.04                     &  -36.38          &     0                       &  5.64                     & 7.04      &  15.4                & 4.1        \\
 &  $\hat h^{(2)}$+ pairing    & 0.0498 & 0.2865 & -36.96             &   -7.04                     &  -42.60          &     0                       &  fit                      & fit      &  14.9                & 4.1       \\
\hline
$^{24}$Mg  &  USD                &  -      &  -      & -73.12             &  -7.00                      &  -80.17          &     0                       &  7.05                     & 7.00      &  18.0                & 6.9         \\
  &  $\hat h^{(2)}$ + pairing    & 0.0254 & 0.4765 & -103.42            &  -7.00                      & -110.47          &     0                       &  fit                      & fit      &  17.5                & 5.7       \\
\hline
$^{36}$Ar  &  USD                &  -      &  -      & -225.29            & -3.55                       & -226.56          &     0                       & 1.27                      & 3.55      & -13.5                & 4.0           \\
  &  $\hat h^{(2)}$ + pairing    & 0.0582 & 0.3554 & -379.57            &   -3.55                     & -380.84          &     0                       & fit                       & fit      &  -12.2               & 2.7       \\
\hline
\hline
\end{tabular}
\end{table}

This leads us to the second energy scale $\delta E_{\rm pair}^{\rm HF+BCS}$ [Eq.~(\ref{delta-E-pair})], which is found to be $7.04, 7.00$ and $3.55$ MeV for $^{20}$Ne, $^{24}$Mg and $^{36}$Ar, respectively. The deformed solution of all three nuclei is found to have no pairing and, as noted in Sec.~\ref{effective-Hamiltonian}, an alternative second energy scale is given by (\ref{spherical-pairing}).

The original space of the USD interaction is already truncated, so orbital truncation effects are not present. The transformation matrix from the deformed to the spherical basis is therefore unitary and the projector ${\bf P}$ is not required. In particular, the consistency relations (\ref{trace-P-rho}) and (\ref{new-conditions}) [or equivalently (\ref{R-proj_first})] are automatically satisfied within the spherical $sd$ shell.

The deformed mean-field Hamiltonian is expressed in the spherical basis [Eq.~(\ref{mf-hamiltonian})] and decomposed into multipoles to find $\hat h^{(0)}$ and $\hat h^{(2)}$. To construct the effective CISM Hamiltonian (\ref{effective-H}), it remains to determine the strength parameters $g_{Q}$ and $g_{P}$ of the quadrupole and pairing interactions. We tune these parameters
to match simultaneously both energy scales (\ref{deformation-energy}) and (\ref{delta-E-pair}) of the USD Hamiltonian when solved in the HF+BCS approximation.
For example, in $^{20}$Ne the USD energy scales of $E_{\rm def}^{\rm HF+BCS}$= 5.64 MeV and
$\delta E_{\rm pair}^{\rm HF+BCS}$=7.04 MeV are matched by similar energies calculated for a CISM Hamiltonian (\ref{effective-H}) with $g_{Q}=0.0498$ MeV/$b^{4}$ and $g_{P}=0.2865$ MeV.  Since the deformed configuration has no pairing solution, we can alternatively match the energy scale (\ref{spherical-pairing}) (together with the deformation energy), resulting in very similar values for the coupling constants $g_Q$ and $g_P$.

To verify the validity of our map, we compare the quadrupole moments of
the mapped CISM Hamiltonian with their mean-field values in the USD interaction. The results, shown in Table \ref{table1}, are in close agreement with each other.

\begin{figure}[h!]
\epsfxsize= 0.6\columnwidth \centerline{\epsffile{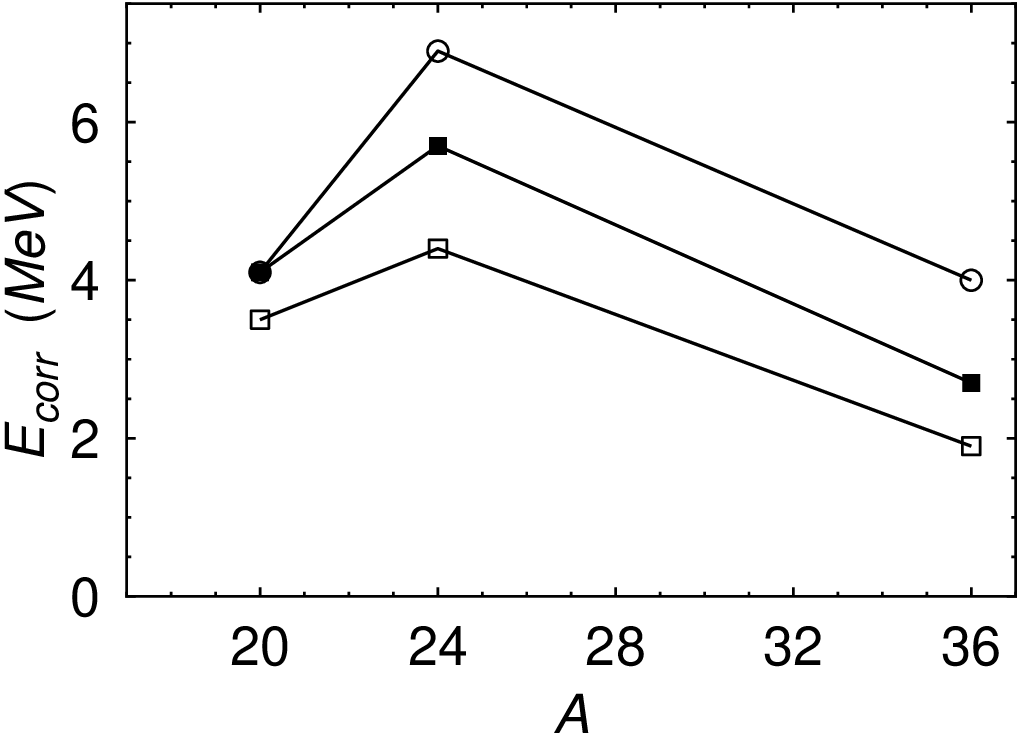}}
\caption{Correlation energies $E_{\rm corr}$ versus  mass number $A$ for the deformed
$N=Z$ $sd$-shell nuclei $^{20}$Ne, $^{24}$Mg and $^{36}$Ar. Results for the mapped CISM Hamiltonian (\ref{effective-H}) (solid squares) are compared with the quadrupolar correlation energies calculated in Ref.~\cite{al06} (open squares) and with the full correlation energies of the USD Hamiltonian (open circles).
}
\label{corr-USD}
\end{figure}

The ground-state energies of the CISM Hamiltonian (\ref{effective-H}) were calculated using
the computer code {\tt oxbash} ~\cite{oxbash} and the corresponding correlation energies are listed in Table \ref{table1}. Figure \ref{corr-USD} shows these correlation energies (solid squares) versus mass number $A$. They are compared with the correlation energies extracted from a mapped CISM Hamiltonian that includes only quadrupolar correlations (open squares)~\cite{al06} and with the full USD correlation energies (open circles). Quadrupole correlations alone seem to be responsible for more than 50\% of the USD correlation energies~\cite{al06}. We observe that the inclusion of pairing correlations in the CISM Hamiltonian (on top of quadrupole correlations) brings us even closer to the full USD correlation energies.  These results are encouraging considering that the USD interaction~\cite{wi84} with 3 single-particle energies 63 anti-symmmetrized interaction matrix elements was fitted to reproduce the observed properties of nuclei in this mass region.

\section{Mapping the DFT}
\label{map-DFT}

In this section we apply the mapping procedure of Sec.~\ref{formalism} starting from a DFT that is based on the SLy4 Skyrme force \cite{ch98} in the particle-hole channel plus a zero-range density-dependent force
\begin{eqnarray} \label{density-dep-pairing}
V({\bf r}_{1},{\bf r}_{2})= - g^{\rm DFT} \left(1-\hat{P}^{\sigma} \right) \left(1-\frac{\rho({\bf r}_{1})}{{\rho}_{c}} \right)
\delta ({\bf r}_{1}-{\bf r}_{2})
\end{eqnarray}
in the pairing channel \cite{te96}. In Eq.~(\ref{density-dep-pairing}), $\hat{P}^{\sigma}$ is the spin exchange operator and $\rho({\bf r})$ the total nuclear density. Throughout this work, we choose the strength of the force $g^{\rm DFT}$ to be the same for protons and neutrons
and the central density is ${\rho}_{c}$=0.16 fm$^{-3}$. A smooth cut-off of 5 MeV around the
Fermi level is used for the interaction (\ref{density-dep-pairing}). The above DFT is solved in the HF+BCS approximation using the Brussels-Paris computer code
{\tt ev8}~\cite{bo85,bo05}.

\subsection{SCMF results}\label{SCMF-results}

\begin{figure}[th!]
\epsfxsize= 0.7 \columnwidth \centerline{\epsffile{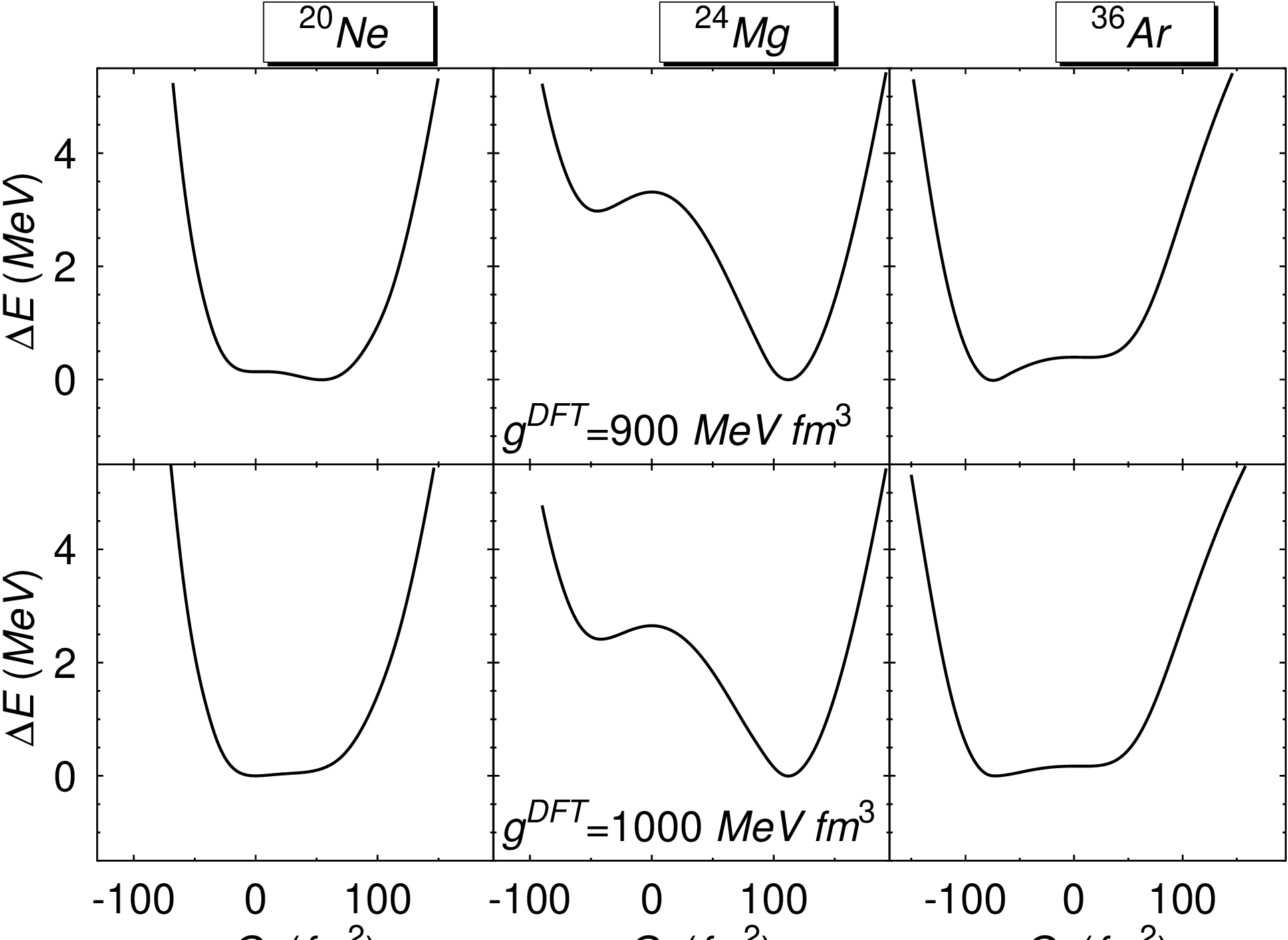}}
\vspace*{7 mm}
\caption{HF+BCS energy curves for the nuclei  $^{20}$Ne (left column), $^{24}$Mg (middle column) and $^{36}$Ar (right column) as functions of the axially symmetric mass quadrupole moment $\langle \hat Q \rangle$. The corresponding
surfaces are shown for DFT pairing
strengths of $g^{\rm DFT}= 900$ and $1000$  MeV-fm$^{3}$. The SLy4 parametrization of the Skyrme interaction is used in the particle-hole channel
while a zero-range density-dependent interaction is included in the pairing channel.
All energies are measured with respect to the global minimum of the corresponding surface.
}
\label{dft-surfaces}
\end{figure}

The starting point of our study is presented in Fig.~\ref{dft-surfaces} where we show HF+BCS energy surfaces versus the axially symmetric mass quadrupole moment $\langle \hat Q \rangle$ for the nuclei  $^{20}$Ne, $^{24}$Mg and $^{36}$Ar. These surfaces are calculated for the SLy4 force plus the zero-range density-dependent force (\ref{density-dep-pairing}) by constraining the value of the mass quadrupole $\langle \hat Q \rangle$. Results are shown for pairing strengths of $g^{\rm DFT}= 900$ and $1000$  MeV-fm$^{3}$. These energy surfaces are useful in finding the global minimum of the HF+BCS energy as well as providing the energy of the spherical solution. Unlike the HF surface of an open-shell nucleus, the $\langle \hat Q \rangle =0$ configuration is always a stationary point of the HF+BCS energy surface.

In the HF approximation, the nucleus $^{20}$Ne is prolate with a deformation
energy of $5.6$ MeV and $\langle \hat Q \rangle_{\rm mf}=84$ fm$^2$~\cite{al06}. In HF+BCS, when including a pairing channel in the DFT force with a strength of $g^{\rm DFT}= 900$ MeV-fm$^{3}$, it remains prolate but the deformation energy and quadrupole moment are reduced to $E_{\rm def}^{\rm HF+BCS}=0.14$ MeV and $\langle \hat Q \rangle_{\rm mf}=57.8$ fm$^2$, respectively. The deformed state is paired with $E_{\rm pair,mf}^{\rm HF+BCS} = -3.3$ MeV. However, $E_{\rm pair}$ is more negative in the spherical state $E_{\rm pair,sph}^{\rm HF+BCS} = -6.16$ MeV, lowering the mean-field energy of the spherical configuration more than in the deformed configuration, and hence the overall reduction of deformation energy. The second energy scale (\ref{delta-E-pair}) that is necessary for the construction of our map is given by $\delta E_{\rm pair}^{\rm HF+BCS} = 2.86$ MeV. The results are tabulated in Table \ref{table2}.

\begin{small}
\begin{table}[h!]
\caption{HF+BCS results for the nuclei $^{20}$Ne, $^{24}$Mg and $^{36}$Ar, using the SLy4 Skyrme interaction plus a zero-range a density-dependent force (\ref{density-dep-pairing}) in the pairing channel, are compared with HF+BCS results for the mapped Hamiltonian (\ref{effective-H}). The coupling constants $g_Q$ and $g_P$ in (\ref{effective-H}) are determined by matching the deformation energy (\ref{deformation-energy}) and pairing energy excess (\ref{delta-E-pair}) of the SCMF theory with those of the CISM. Energies are in MeV, $ \langle \hat Q \rangle$ is in fm$^2$ and $g_{Q}$ is in MeV/fm$^{4}$. For the mapped Hamiltonian $g_P$ is in MeV while for the SCMF $g_{P} \equiv g^{\rm DFT}$ is in MeV-fm$^{3}$. The last column shows the correlation energy (\ref{E-corr}), calculated from the ground-state energy of the CISM Hamiltonian with the coupling parameters shown in the Table.
}
\label{table2}
\vspace*{1 mm}
\begin{tabular}{|r|c|c|c|c|c|c|c|c|c|c|c|c|}
\hline
\hline
Nucleus  &  Interaction             & $g_{Q}$ & $g_{P}$              & $E_{\rm sph}^{\rm HF+BCS}$ & $E_{\rm pair,sph}^{\rm HF+BCS}$ & $E_{mf}^{\rm HF+BCS}$ & $E_{\rm pair,mf}^{\rm HF+BCS}$ & $E_{\rm def}^{\rm HF+BCS}$ & $\delta E_{\rm pair}^{\rm HF+BCS}$ & $ \langle \hat Q \rangle$& ${\langle \hat Q \rangle}_{mf}$ & $E_{\rm corr}$\\
\hline
$^{20}$Ne  &  SLy4               &  -      & 900                  & -157.91            &  -6.16           &  -158.05       &  -3.30             &  0.14            &   2.86 &
34.8  &57.8  &  -\\
  &  $\hat h^{(2)}$ + pairing    & 0.0954 & 0.4516              & -35.58             &  -5.00          &  -35.72        &  -2.14             &  fit             &   fit    &
  35.7     &   -                            &  3.4  \\
\hline
\hline
$^{24}$Mg  &  SLy4               &  -      & 900                  & -192.61             & -5.84              &  -195.93        & 0            &  3.32                & 5.84      &   59.4  & 111.9           &  -\\
    &  $\hat h^{(2)}$ + pairing    & 0.0343 &  0.4930             &  -85.12             & -5.84              &  -88.44         & 0            &  fit                & fit      &
 58.4      &   -                            &  4.7  \\
\hline
$^{24}$Mg  &  SLy4                &  -     &  1000                & -193.27             & -6.67              &  -195.93        &  0           &  2.66                & 6.67      &    59.4                 &  111.9                  &  -      \\
      &  $\hat h^{(2)}$ + pairing     & 0.0332 &  0.5729            & -86.02              & -6.67              &  -88.68         & 0            &  fit                & fit      &    58.0                 &   -                            &
5.1 \\
\hline
\hline
$^{36}$Ar  &  SLy4               &  -      & 900                  & -305.02             & -2.14              & -305.43         &  0           &  0.41            & 2.14                &  -43.4
& -74.7            & - \\
      &  $\hat h^{(2)}$+ pairing    & 0.0992 &  0.3128             & -281.15             & -2.14              & -281.56         & 0            &  fit            &  fit               &   -38.8   &   -
&  2.0  \\
\hline
$^{36}$Ar  &  SLy4               &  -      &  1000                & -305.26             &  -2.48            & -305.43          &  -0.53          & 0.17             & 1.95                &  -39.7              & -68.9
 & -\\
      &  $\hat h^{(2)}$ + pairing     & 0.1017 & 0.3275             & -283.15             & -1.95             &  -283.32         &  0            &  fit            & fit                &   -35.4   &     -
&  1.9  \\
\hline
\hline
\end{tabular}
\end{table}
\end{small}

As $g^{\rm DFT}$ increases beyond $900$ MeV-fm$^{3}$, a shape transition occurs in $^{20}$Ne to a spherical shape, and for $g^{\rm DFT}= 1000$ MeV-fm$^{3}$ the ground-state configuration is spherical. In the present work, we discuss the mapping for nuclei whose mean-field solution is deformed, so the case $g^{\rm DFT}=1000$ MeV-fm$^{3}$ is excluded for $^{20}$Ne.

The nucleus $^{24}$Mg is also prolate in the HF approximation with a deformation energy
of $8.73$ MeV and  $\langle \hat Q \rangle_{\rm mf}=111.9$ fm$^2$~\cite{al06}. When a zero-range density-dependent force is included in the DFT pairing channel, $^{24}$Mg  remains prolate with the same deformation of $\langle \hat Q \rangle_{\rm mf}=111.9$ fm$^2$ for both values of $g^{\rm DFT}$. This is since the deformed HF+BCS ground state is found to have $\kappab=0$. However, the spherical HF+BCS energy is lowered by the pairing force, decreasing the deformation energy to $E_{\rm def}^{\rm HF+BCS}=3.32$ and $2.66$ MeV for $g^{\rm DFT}= 900$ and $1000$  MeV-fm$^{3}$, respectively.  In these cases, the pairing energy excess (\ref{delta-E-pair}) is the absolute value of the spherical pairing energy and is observed to increase for larger values of the DFT pairing strength $g^{\rm DFT}$ (see Table \ref{table2}).

Finally, the nucleus $^{36}$Ar is oblate in the HF approximation with a deformation energy of $2.27$ MeV and $\langle \hat Q \rangle_{\rm mf}=-74.7$ fm$^2$~\cite{al06}, and it remains oblate with the same deformation in HF+BCS for $g^{\rm DFT}=900$ MeV-fm$^{3}$ (no pairing in the deformed solution). The pairing energy of the spherical configuration increases slightly with $g^{\rm DFT}$. However, the deformed solution becomes weakly paired for $g^{\rm DFT}= 1000$ MeV-fm$^{3}$, and the ground-state quadrupole moment is reduced in magnitude to $\langle \hat Q \rangle_{\rm mf}=-68.9$ fm$^2$. Consequently, we observe a small decrease in the value of $\delta E_{\rm pair}^{\rm HF+BCS}$ (see Table \ref{table2}).

The DFT calculations provide us with the single-particle mean-field Hamiltonian of the deformed configuration, necessary for the construction of $\hat h^{(0)}$ and $\hat h^{(2)}$ in the effective CISM Hamiltonian (\ref{effective-H}), as well as with the two energy
scales $E_{\rm def}^{\rm HF+BCS}$ and $\delta E_{\rm pair}^{\rm HF+BCS}$ required for the determination of the two coupling parameters $g_Q$ and $g_P$ in (\ref{effective-H}).

\subsection{Truncation effects and operator rescaling}\label{truncation}

The next step in constructing the map involves a transformation to the spherical basis and truncation of the spherical shell model space to the valence shells. This truncation was absent in our USD studies in Sec.~\ref{USD}, since the original space was already truncated.  Here the situation is different since the DFT includes a large number of spherical orbitals and it is necessary to truncate to the spherical $sd$ shell. Thus, before we proceed with the construction of the CISM Hamiltonian, we study in this section truncation effects on the density matrix and pairing tensor, as well as on the quadrupole moment.

The HF+BCS calculations determine the density matrix ${\rhob}$ and
pairing tensor $\kappab$ (separately for protons and neutrons) of the deformed
mean-field ground state. We transform these matrices to the spherical DFT basis using Eqs.~(\ref{trans-mat}) and, as long as this spherical basis is complete, they continue to satisfy the consistency conditions (\ref{trace-rho}) and (\ref{new-conditions}).

Once the model space is truncated, the matrices ${\rhob}$ and $\kappab$ are projected on the valence $sd$ shell to give ${\bf P}{\rhob}{\bf P}$ and ${\bf P}{\kappab}{\bf P}$. We expect the mapping procedure to work well when the projected density matrix and pairing tensor satisfy approximately the consistency relations (\ref{trace-P-rho}) and (\ref{new-conditions}).

When the deformed HF+BCS solution is unpaired, $\kappab=0$ and relations (\ref{new-conditions}) reduce to the usual HF consistency relation $\rhob^2=\rhob$. For such cases, this relation together with (\ref{trace-P-rho}) were already found to hold to a very good approximation in Ref.~\onlinecite{al06}. Hence, we need to study truncation effects only in cases where the deformed mean-field solution is paired ($\kappab\neq 0$).  These cases are $^{20}$Ne for $g^{\rm DFT}=900$ MeV-fm$^3$ and $^{36}$Ar for $g^{\rm DFT}=1000$ MeV-fm$^3$ (see Table \ref{table2}).

Rather than studying relations (\ref{new-conditions}), it is more
convenient to study the equivalent requirement that the projected generalized
density  matrix $\calRb$ is a projector in twice the number of
dimensions [see Eq.~(\ref{R-proj_first})]. Relation (\ref{R-proj_first}) is satisfied exactly when
$\calRb$ has eigenvalues that are either $1$ or $0$ and trace that is twice the number of valence single-particle orbitals. Table \ref{table3} lists the eigenvalues and trace of the matrix ${\bf P} \calRb {\bf P}$. The dimension of ${\bf P} \calRb {\bf P}$ in the $sd$ shell is 24 for
each type of nucleon. However, the eigenvalues of ${\bf P} \calRb {\bf P}$ are pairwise
degenerate (because of time-reversal symmetry) and we list only the 12 distinct
eigenvalues (for either protons or neutrons). We observe that all eigenvalues are very close
 to either $1$ or $0$, so that the projected generalized density matrix satisfies the HF+BCS consistency condition $\calRb^2=\calRb$ to a good accuracy.

\begin{table}[h!]
\caption{Eigenvalues and trace of the truncated generalized density matrix
${\bf P}\calRb {\bf P}$ of the deformed ground state in the SCMF theory of the SLy4 Skyrme
energy functional plus a zero-range density-dependent force. The last column shows the trace of the truncated density matrix ${\bf P}\rhob{\bf P}$. Results are shown for those cases in Table \ref{table2} that have a paired deformed solution ($\kappab \neq 0$).
}
\label{table3}
\vspace*{1 mm}
\begin{tabular}{|c|c|c|c|cccccc|c|c|c|}
\hline
\hline
Nucleus  &  Interaction  &  $g^{\rm DFT}$  & nucleon      &  & &  & Eigenvalues of ${\bf P}\calRb {\bf P}$ &  &  & $tr\!\!\left({\bf P}\calRb {\bf P}\right)$ & $tr\!\!\left({\bf P}\!\rhob {\bf P}\right)$ \\
\hline
$^{20}$Ne  &  SLy4  & 900     & neutron          & $-6 \times 10^{-6}$ & $-1 \times 10^{-6}$ & $4 \times 10^{-6}$ & $2.6 \times 10^{-5}$  & $ 1.7 \times 10^{-4}$ & 0.003  & 23.824  & 1.986   \\
          &          &         &                  &  0.986 &  0.991 & 0.991  & 0.992 & 0.995& 0.996  &    &   \\
    &     &     & proton           & $-3\times 10^{-6}$ & $1\times 10^{-6}$ & $ 5\times 10^{-6}$  & $ 7.1 \times 10^{-5}$ & $3.8 \times 10^{-4}$ & 0.003  & 23.774   & 1.985 \\
           &          &         &                  &  0.983 &  0.985& 0.991  & 0.991 & 0.995  & 0.996 &    &   \\
\hline
\hline
$^{36}$Ar  &  SLy4  & 1000     & neutron         & $-2 \times 10^{-6}$ & $1.1 \times 10^{-5}$  & $4.1\times 10^{-5}$ & $8.2 \times 10^{-5}$ & $1.9 \times 10^{-4}$ & $8 \times 10^{-4}$ & 23.918 & 9.964  \\
           &          &         &                  &  0.993 & 0.995  & 0.996  & 0.997  & 0.998 & 0.999  &    &   \\
   &       &    & proton          & $-1 \times 10^{-5}$ & $-5 \times 10^{-6}$  & 0  & $8\times 10^{-6}$  & $2.3 \times 10^{-5}$  & $8.5 \times 10^{-4}$  &  23.914   & 9.956 \\

   &          &         &                  &  0.993 & 0.995  & 0.996 & 0.997 & 0.998  & 0.999 &   &    \\
\hline
\hline
\end{tabular}
\end{table}

Another condition is (\ref{trace-P-rho}) (i.e., the trace of the projected density ${\bf P} \rhob {\bf P}$ should be approximately equal to the valence number of nucleons of each type). When $\kappab=0$, this condition was already found to be well satisfied in Ref.~\onlinecite{al06}. For cases with $\kappab \neq 0$, the respective traces are tabulated in the last column of Table \ref{table3}. We observe that condition (\ref{trace-P-rho}) is satisfied to a good accuracy also when the ground-state solution is paired. We note that while the number of particles is fixed in the HF approximation, it is only the average number of particles that is fixed in the HF+BCS approximation.  We also note that when the deformed ground state has $\kappab \neq 0$, the eigenvalues of $\rhob$ correspond to the BCS parameters $v_k^2$ and they can differ significantly from both 1 and 0.

The SCMF theory allows us to determine the rescaling of one-body observables in the CISM model space by comparing their expectation values in the complete and truncated spaces. A good
example is the mass quadrupole operator. In Table \ref{table2} we compare
its expectation value $\lb \hat Q\rb_{\rm mf}$ in the complete space with its expectation  value $\lb\hat Q\rb$ in the truncated space ($sd$ shell). The ratio $\lb \hat Q\rb_{\rm mf}/\lb \hat Q\rb$ determines the rescaling factor of the mass quadrupole moment when working in the truncated CISM space. In cases where $\kappab =0$, these scaling factors coincide with those found in the HF theory~\cite{al06}. An interesting issue is the dependence of the rescaling factor on the strength $g^{\rm DFT}$  of the DFT zero-range density-dependent force, once the latter is sufficiently strong to induce pairing in the deformed ground-state solution. For $^{20}$Ne, we find that the quadrupole operator should be rescaled by  $1.66$ for $g^{\rm DFT}= 900$ MeV-fm$^3$, compared with $1.71$ in the absence of a pairing channel in the DFT. This is a slight reduction of the rescaling factor by about $2.3\%$, even though the quadrupole moment itself is reduced by about $31\%$. Thus, the rescaling of the mass quadrupole operator in the truncated CISM space seems to be almost independent of $g^{\rm DFT}$.

\subsection{CISM Hamiltonian}\label{mapped-H}

Once the monopole and quadrupole parts ($\hat h^{(0)}$ and $\hat h^{(2)}$) of the projected
mean-field Hamiltonian are determined, we can construct the effective
CISM Hamiltonian (\ref{effective-H}) except for the two unknown coupling
constants $g_Q$ and $g_P$. As already discussed in Sec.~\ref{effective-Hamiltonian}, these coupling constants are determined by matching simultaneously the DFT energy scales $E_{\rm def}^{\rm HF+BCS}$ and $\delta E_{\rm pair}^{\rm HF+BCS}$ with similar energy scales for the CISM Hamiltonian when the CISM is solved in the HF+BCS approximation. The values of $g_Q$ and $g_P$ for the various cases studied here are listed in Table \ref{table2}.
 In general, the spherical and deformed pairing energies of the
mapped CISM Hamiltonian differ from the corresponding DFT pairing energies. For example, in
$^{20}$Ne with $g^{\rm DFT}= 900$ MeV-fm$^3$, we find the values of $E_{\rm pair,sph}^{\rm HF+BCS}= -5.00$ MeV and $E_{\rm pair,mf}^{\rm HF+BCS}= -2.14$ MeV in a mean-field solution of the CISM Hamiltonian, while the DFT values are $E_{\rm pair,sph}^{\rm HF+BCS}= -6.16$ MeV and $E_{\rm pair,mf}^{\rm HF+BCS}= -3.30$ MeV (see Table \ref{table2}). It is only the difference $\delta E_{\rm pair}^{\rm HF+BCS}$ in pairing energies between the deformed and spherical solutions that is matched ($2.86$ MeV in this case).

We next study how well can the CISM Hamiltonian reproduce one-body observables such as the mass quadrupole. In Table \ref{table2} we compare the value of $\lb \hat Q\rb$ calculated when the CISM Hamiltonian (\ref{effective-H}) is solved in the mean field (i.e., HF+BCS) approximation, with its SCMF value in the truncated space ($sd$ shell).  For $^{20}$Ne at $g^{\rm DFT} = 900$ MeV-fm$^3$, we find the CISM value of $\lb \hat Q\rb =35.7$ fm$^2$ as compared with the DFT value of $\lb \hat Q\rb =34.8$ fm$^2$ in the truncated $sd$ space. We find a similarly close agreement for $^{24}$Mg. The largest deviation (about $11\%$) is found for $^{36}$Ar.

\begin{table}[b!]
\caption{Neutron and proton spherical occupations $\lb n_{j}\rb$ for the nuclei $^{20}$Ne, $^{24}$Mg and $^{36}$Ar in the $sd$-shell valence space. The DFT (SLy4 with a zero-range density-dependent force in the pairing channel) results are compared with the occupations of the mapped CISM Hamiltonian (\ref{effective-H}) both in the HF+BCS approximation and in shell model calculations.  For the SLy4 interaction $g_P \equiv g^{\rm DFT}$.
}
\label{table4}
\vspace*{1 mm}
\begin{tabular}{|c|c|c|c|c|c|c|c|c|c|}
\hline
\hline
Nucleus  &  Interaction                   & $g_Q$ & $g_P$       & $n0d5/2$ & $n1s1/2$ & $n0d3/2$  & $p0d5/2$ & $p1s1/2$ & $p0d3/2$ \\
\hline
$^{20}$Ne  &  SLy4                       & -       & 900           & 1.713    & 0.221    & 0.052     & 1.709  & 0.225    & 0.052     \\
        & $\hat h^{(2)}$ + pairing (HF+BCS)     & 0.0954 & 0.4516       & 1.736    & 0.216    & 0.048     & 1.731  & 0.222    & 0.048    \\
        & $\hat h^{(2)}$ + pairing (CISM)       & 0.0954 & 0.4516       & 1.730  & 0.226    & 0.045     & 1.724  & 0.230    & 0.046    \\
\hline
\hline
$^{24}$Mg  &  SLy4                       & -       & 900           & 3.293    & 0.408    & 0.215     & 3.281    & 0.401    & 0.215         \\
         & $\hat h^{(2)}$ + pairing (HF+BCS)     & 0.0343 &  0.4930      & 3.489    & 0.304    & 0.207     & 3.487    & 0.306    & 0.207        \\
      & $\hat h^{(2)}$ + pairing (CISM)       & 0.0343 &  0.4930      & 3.441    & 0.299    & 0.260     & 3.435    & 0.303    & 0.263       \\
\hline
\hline
$^{36}$Ar  &  SLy4                       & -       & 1000          & 5.934    & 1.793    & 2.237     & 5.930   & 1.786    & 2.239       \\
        & $\hat h^{(2)}$ + pairing (HF+BCS)     & 0.1017 &  0.3275       & 5.995    & 1.888    & 2.117     & 5.995    & 1.884   &  2.122    \\
        & $\hat h^{(2)}$ +pairing (CISM)       & 0.1017 & 0.3275       &  5.976   & 1.836    & 2.188     &  5.975   & 1.831    &  2.194    \\
\hline
\hline
\end{tabular}
\end{table}

\begin{figure}[t!]
\epsfxsize= 0.9 \columnwidth \centerline{\epsffile{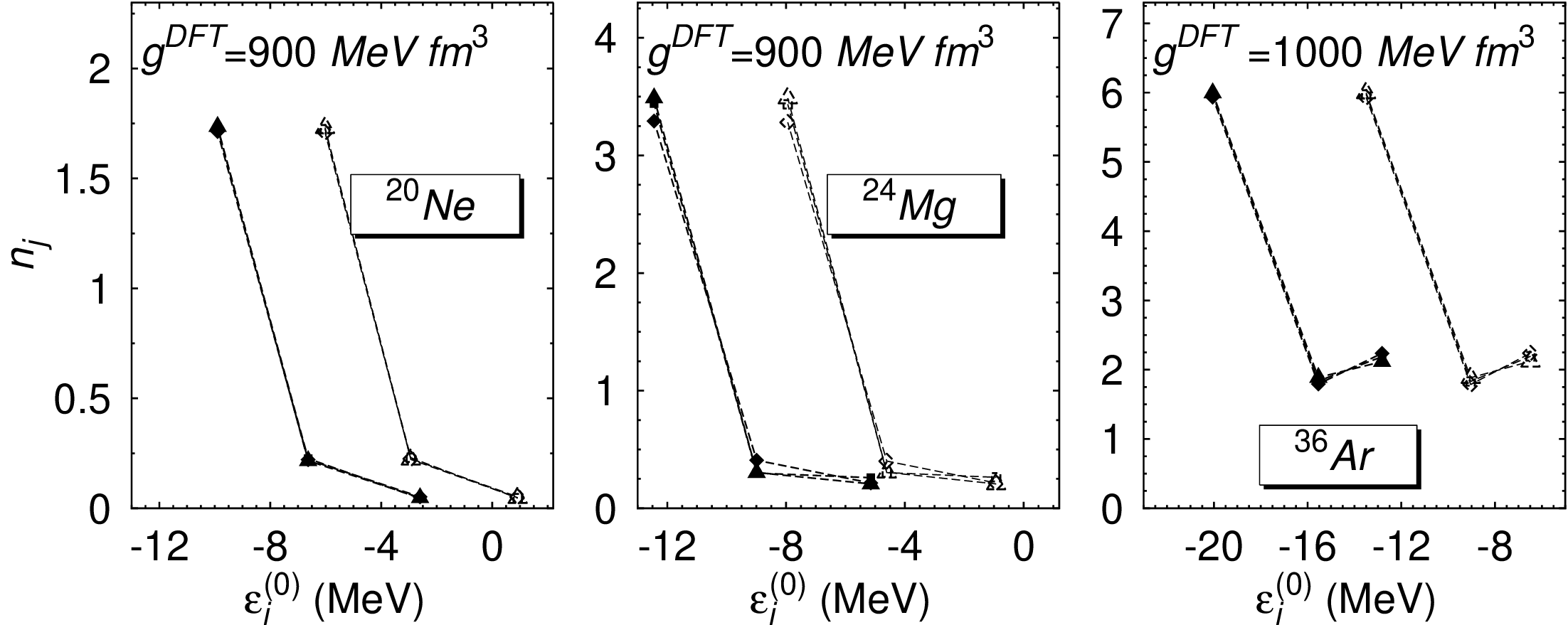}}
\caption{The occupations $\lb \hat n_j \rb$ of the valence spherical
orbitals as a function of the  single-particle energy $\epsilon^{(0)}_j$ for the nuclei $^{20}$Ne, $^{24}$Mg and $^{36}$Ar (see Table \ref{table4}). The DFT results (diamonds) are compared with the results in the HF+BCS ground-state solution of the CISM Hamiltonian (\ref{effective-H}) (triangles). Also shown are the occupations calculated in a shell model approach for the effective CISM Hamiltonian (squares).  Open (solid) symbols are for protons (neutrons).
}
\label{occupations}
\end{figure}

Other relevant one-body quantities are the occupations $\lb \hat n_j\rb$ of the spherical orbitals. In Table \ref{table4} we compare their DFT values $\lb \hat n_j\rb=\sum_{m} \rho_{jm,jm}$ with the respective spherical occupations of the CISM when solved in the HF+BCS approximation. We also list in Table \ref{table4} the CISM occupations as calculated by the shell model code {\tt oxbash}.  Results are shown for $^{20}$Ne and $^{24}$Mg at $g^{\rm DFT}=900$ MeV-fm$^3$, and $^{36}$Ar at $g^{\rm DFT}=1000$ MeV-fm$^3$. Even in cases where the deformed ground state is unpaired, the CISM occupations are expected to differ from their values in Ref.~\onlinecite{al06}. The reason is that the CISM Hamiltonian studied here contains an additional pairing term.  The occupations computed in all three methods are in rather close agreement.  These various spherical occupations are also shown in Fig.~\ref{occupations} versus the corresponding spherical single-particle energy ${\epsilon}^{(0)}_{j}$.

\begin{figure}[b!]
\epsfxsize= 0.6 \columnwidth \centerline{\epsffile{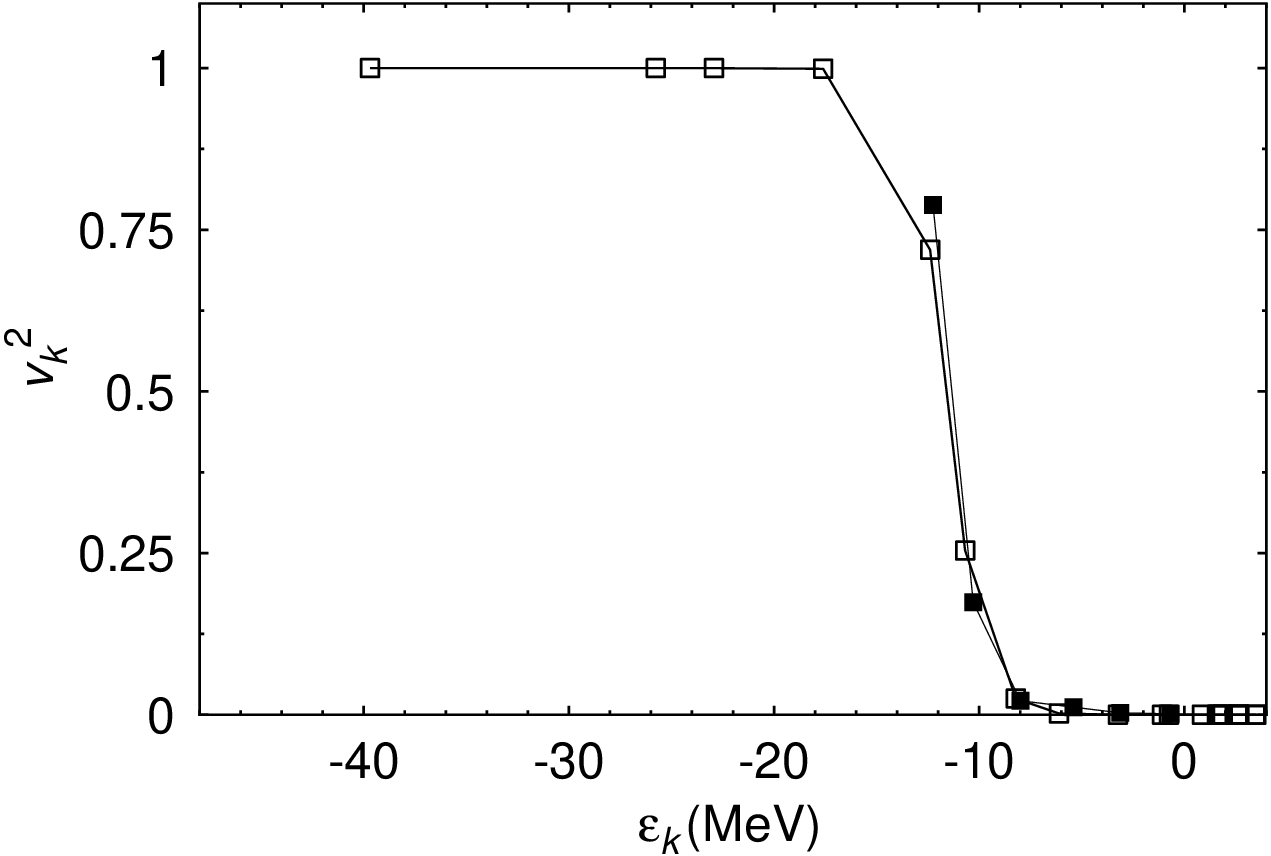}}
\caption{Neutron BCS occupations $v_k^2$ of the deformed ground-state solution of the CISM Hamiltonian (\ref{effective-H}) (solid squares) are compared with the BCS occupations in the deformed basis of the DFT ground-state solution (open squares). These BCS occupations are plotted versus the deformed single-particle HF energies $\epsilon_k$. Results are shown for $^{20}$Ne using the SLy4 parametrization of the Skyrme force and a zero-range density-dependent force with $g^{\rm DFT} = 900$ MeV-fm$^3$. }
\label{v2-BCS}
\end{figure}

Finally, we compare the BCS occupations $v_k^2$ of the deformed solution of the mapped CISM Hamiltonian (defined in the truncated space) with the $v_k^2$ of the original SCMF ground state (defined in the complete space). Figure \ref{v2-BCS} makes this comparison for $^{20}$Ne at $g^{\rm DFT} = 900$ MeV-fm$^3$. The $v_k^2$ of the mapped theory are slightly enhanced (suppressed) below (above) the Fermi energy when compared with their DFT values. In the DFT, core orbitals below the $sd$ shell have $v_k^2 \approx 1$, while orbitals above the valence $sd$ shell have $v_k^2 \approx 0$.

\subsection{Correlation energies}

\begin{figure}[h!]
\epsfxsize= 0.57 \columnwidth \centerline{\epsffile{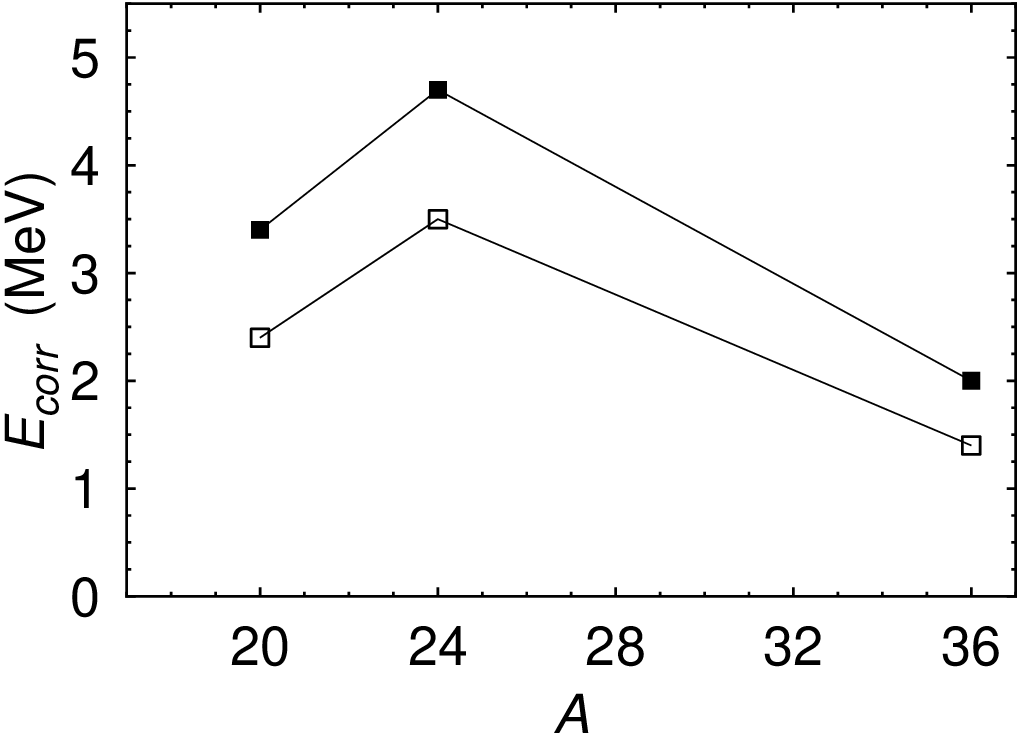}}
\caption{Correlation energies $E_{\rm corr}$ [see Eq.~\ref{E-corr})] versus mass number A for the
$N=Z$ $sd$-shell nuclei $^{20}$Ne, $^{24}$Mg and $^{36}$Ar, calculated using the mapped CISM Hamiltonian (\ref{effective-H}) (solid squares). The SCMF theory used to construct the map is the SLy4 Skyrme force plus a zero-range density-dependent force in the pairing channel of strength $g^{\rm DFT}= 900$ MeV fm$^{3}$. The open squares are the correlation energies found from a mapped CISM Hamiltonian that includes quadrupolar correlations alone~\cite{al06}.
}
\label{corr-DFT}
\end{figure}

An important application of the DFT to CISM map is the calculation of correlation energies. We calculate the ground-state energy $E_{\rm gs}$ of the mapped Hamiltonian (\ref{effective-H}) in the shell model approach and determine the correlation energy using Eq.~(\ref{E-corr}). The correlation energies for the three nuclei $^{20}$Ne, $^{24}$Mg and  $^{36}$Ar are tabulated in Table \ref{table2}. In general, the correlation energy of the mapped Hamiltonian is found to increase with $g^{\rm DFT}$. In Fig.~\ref{corr-DFT} we show the correlation energy versus mass number $A$ for  $g^{\rm DFT}=900$ MeV fm$^{3}$.  We compare our results (solid squares) with the correlation energies obtained in Ref.~\onlinecite{al06} where the effective interaction includes only a quadrupolar component (open squares).  The inclusion of pairing correlations provides an additional gain in correlation energy and the observed trend versus mass number $A$ is similar to the trend observed in the USD results shown in Fig.~\ref{corr-USD}.

Our correlation energies of $5.1$ MeV and $1.9$ MeV for the nuclei $^{24}$Mg and  $^{36}$Ar, calculated for $g^{\rm DFT}=1000$ MeV fm$^{3}$, are in very close agreement with the values of $5.1$ MeV and $2.0$ MeV obtained in Ref.~\onlinecite{be06} starting from the same DFT but using an angular-momentum projected GCM. The correlation energy of $3.4$ MeV we find for $^{20}$Ne (using a somewhat reduced value of $g^{\rm DFT}$) is also consistent with the value of 3.1 MeV reported in Ref.~\onlinecite{be06}.

\section{Conclusion} \label{conclusion}

In this work we have mapped an SCMF theory onto an effective CISM Hamiltonian in the presence of both quadrupolar and pairing correlations. When compared with Ref.~\onlinecite{al06}, this is an important step forward in constructing a general map that would take advantage of the global validity of the energy density functional and the higher accuracy of the CISM approach. The main differences compared with the study initiated in Ref.~\cite{al06} are: (i) The SCMF is now based on an HF+BCS approximation (rather than HF), and the ground-state consistency conditions are recast in terms of both the density matrix and the pairing tensor. (ii) The effective Hamiltonian depends now on two coupling strengths, and consequently two energy scales must be matched to determine these coupling parameters. One energy scale is the deformation energy, similar to the one used previously but now redefined in terms of the HF+BCS energies. The second scale is determined by the difference of average pairing energy in the deformed ground state and spherical configurations. The new map constructed here provides improved correlation energies that are consistent with other methods that include quadrupolar and pairing correlations beyond the mean field.

The DFT application discussed here is based on a Skyrme force in the particle-hole channel plus a zero-range density-dependent interaction in the pairing channel. One could alternatively start from a mean-field calculation with a finite-range Gogny force. Such an interaction includes pairing correlations consistently and is usually treated in the  Hartree-Fock-Bogolyubov (HFB) method. The map constructed here can be easily generalized to the Gogny DFT by replacing the HF+BCS approximation by the HFB approximation, both when solving the DFT and in the mean-field solution of the CISM Hamiltonian.

As already discussed in Ref.~\cite{al06} much remains to be done. Nuclei that have no spherical pairing solution can be treated by introducing a constraining pairing field. The addition of an interaction term built from an isovector quadrupole operator would enable us to generalize our studies to $N \neq Z$ nuclei and carry out systematic studies. A major challenge is the application to the heavy deformed nuclei. Although the method presented here can in principle be adapted to such nuclei, outstanding issues are the ability of a spherical shell model to describe nuclei with large deformation in a truncated space and the practical issue of solving the CISM Hamiltonian in large model spaces.

\begin{acknowledgments}
This work was supported in part by the U.S. DOE grants No. DE-FG02-91ER40608 and FG02-00ER41132.
\end{acknowledgments}

\end{document}